\documentclass[preprint,pre]{revtex4}
\usepackage{amssymb}
\usepackage{amsmath}
\usepackage{epsfig}
\usepackage{enumerate}

\def\bra#1{\mathinner{\langle{#1}|}}
\def\ket#1{\mathinner{|{#1}\rangle}}

\def\leer{\varnothing}

\def\id3{ {\mathbf{1}_{3\times 3}}}

\def\0{\varnothing}
\def\1{{\bf 1}}
\begin{document}

\title{Reaction fronts in stochastic exclusion models with three--site interactions}
\author{Matthias Paessens}
\email{m.paessens@fz-juelich.de}
\author{Gunter M. Sch\"utz}
\affiliation{Institut f\"ur Festk\"orperforschung, Forschungszentrum J\"ulich - 
52425 J\"ulich, Germany}
\date{\today}

\begin{abstract}
{
The microscopic structure and movement of reaction fronts in reaction diffusion systems 
far from equilibrium are investigated. 
We show that some three--site interaction models exhibit exact diffusive
shock
measures, i.e. domains of different densities connected by a sharp wall without correlations. 
In all cases fluctuating domains grow at the expense of ordered domains, the absence of
growth is possible between ordered domains.
It is shown that these models give rise to aspects not seen in nearest neighbor models,
viz. double shocks and additional symmetries. A classification of the systems
by their symmetries is given and the link of domain wall motion and a free fermion description 
is discussed.  
}

\end{abstract}

\maketitle
\section{Introduction}
The emergence of patterns and fronts is a challenging problem in biology, chemistry and
physics, for a review see Ref.~\cite{Cros94}. In biology for example
bacteria aggregate building up regions with a high density in coexistence
with regions with a low density of organisms \cite{Welc01}. A typical
example in chemistry is the movement of reaction fronts.
In physics the movement of domain walls is directly related to the problem
of coarsening, for example in magnetic systems. In Ref.~\cite{Rako03}
the phenomenon of hysteresis in a driven diffusive system is
explained by the movement of shocks, i.e., a jump in the density profile.
Various other phenomena in many particle systems can be attributed to the emergence
of shocks, for example the first order transition in the 
phase diagram of the asymmetric exclusion process (ASEP) \cite{Schu01} or phase 
coexistence in a driven diffusive
system coupled to reaction kinetics \cite{Parm03,Popk03}. 
Recently, shocks in quantum systems have also been discussed \cite{Huny03,Popk00}.

On a macroscopic level shock fronts are described by partial differential 
equations. The most prominent equations are the Fisher and the Burgers'
equations. The Fisher equation \cite{Fish37}
\begin{equation}
\frac{\partial}{\partial t} \rho(x,t) = D \frac{\partial^2}{\partial x^2} \rho(x,t) + a \rho(x,t)
-b \rho^2(x,t) 
\end{equation}
was originally proposed as a model for the propagation of a mutant gene. It shows traveling wave
solutions and may be used for modeling systems without conservation of the order
parameter. 

The inviscid Burgers' equation \cite{Burg74}
\begin{equation}
\frac{\partial}{\partial t} \rho(x,t) = - a \frac{\partial}{\partial x} \left[\rho(x,t)\left(1-\rho(x,t)\right)\right] 
\end{equation} 
was proposed as a model for turbulent fluid motion. It as well shows shock solutions, but it may
be used for modeling systems with particle conservation.

In this paper we demonstrate for some models 
how these macroscopic shocks originate from the 
microscopic dynamics. It is known that some driven diffusive systems can be described by the
Fisher or Burgers' equation in the hydrodynamic limit. This limit is achieved by scaling the
lattice constant to zero while keeping the overall length of the system constant, the time
has to be rescaled appropriately. One of these models is the ASEP which is in the hydrodynamic
limit described by the Burgers' equation. Another model is the branching and coalescing
random walk, which is in the hydrodynamic limit described by the Fisher equation \cite{benA98}.
In these two models the microscopic dynamics can be described by exact shock measures.
An exact shock measure is a state where two product measures 
with different densities
are connected. The time evolution of this state is given by a diffusion equation with 
respect to the
position of the density step \cite{Pigo00,Beli02,Kreb03}. In particular this
implies that the microscopic structure of the system is known
at all times.

The physical properties of large classes of one--species reaction--diffusion models with
{\em nearest neighbor} interactions have been widely studied \cite{Schu01,Schu95,Alim01,Mobi01}.
There are only four known 
models which show shocks without correlations \cite{Kreb03,Bala01,Bala04}:
the ASEP, the 
branching and coalescing
random walk (BCRW), the asymmetric Kawasaki--Glauber process (AKGP)
and  the brick layer model. While the former three models are exclusion
models where the number of particles on each lattice site is
restricted to at most one, the particle number is not restricted in the
latter one. 
Here, the investigation shall be extended to three--site interaction exclusion 
models \cite{Khor03,Henk01} and it will be shown 
that three--site interactions 
give rise
to models with exact shock measures which show aspects not seen 
in nearest neighbor models, viz. 
double shocks and additional symmetries even though no free fermion condition is met.

At this point we remark that
the question of phase separation is directly linked to the movement
of domain walls because coarsening is generic for this phenomenon. But although 
some of the models presented in this paper show growing domains we will argue that
this mechanism cannot be used for constructing nonequilibrium models with
two species (one type of particles and vacancies) 
showing phase separation in one dimension \cite{Gray01,Gacs01,Cues03}.    

We present a survey of one--species models with three--site 
interactions and open boundaries
whose time evolution is described by an exact diffusive shock measure, 
to be defined below.
For the cases of shocks between two nonfluctuating phases (densities 0 to 1) 
and two fluctuating phases (both densities are different from 0
and 1) the list is complete. For the case of a shock from a 
nonfluctuating phase (density 0 or 1) to 
a fluctuating phase (density between 0 and 1) the variety of
models is too large to give a complete survey -- the number 
of free parameters rises from 12 to 56 when the interaction
range is increased from two to three. But we present a 
classification of models with respect to their
symmetries, where we considered models where at least two of the 
symmetries charge ($C$), parity ($P$) and time ($T$) are valid
independently. 

We also address the question to what extent the description of the dynamics of
the model by the movement of shock fronts is sufficient. To this end the interactions
of the domain walls are determined and possible parallels to free fermion systems
are discussed.

\section{Formalism}
On each lattice site $k$
($k=1, \ldots, L$) of a one--dimensional lattice there
may be at most one particle ($A$) or a vacancy ($\0$).
One can also consider these two--state systems as spin systems, a particle
is represented by a down spin ($\ket{\downarrow}$)
and a vacancy by an up spin ($\ket{\uparrow}$).
The stochastic dynamics of the models are defined by a master equation
\cite{Kamp01} which is conveniently expressed in the quantum--Hamiltonian 
formalism for spin--1/2 chains 
as described in Ref.~\cite{Schu01}. 
To each lattice 
configuration $\eta$ we assign a basis vector
$\ket{\eta}$ which is given by the tensor product of the 
single--site states. The probability vector
describing the system can then be written as
\begin{equation}
\ket{P(t)}=\sum_{\eta} P(\eta,t) \ket{\eta}
\end{equation}
where $P(\eta,t)$ is the probability at time $t$ to find the system in the state $\eta$. 
The time evolution of the system is described by a master equation which can be written as
\begin{equation}
\frac{d}{dt} \ket{P(t)}=-H\ket{P(t)}
\label{3PWW:MG}
\end{equation}
where $H$ is the stochastic generator of the process. Due to the analogy of equation
(\ref{3PWW:MG}) to the Schr\"odinger equation (in imaginary time) $H$ is often
called the Hamiltonian of the system.
Conservation of probability requires that the sum of the entries of each column is
zero,
\begin{equation}
\bra{s} H = 0,
\end{equation}
where $\bra{s}=(1,1,...)$ is the so-called summation vector.

For finite interaction range and spatially homogeneous
kinetics it is convenient to write the generator as
\begin{equation}
H=-\sum_k h_k-b_1 - b_{L-1},
\end{equation}
where the local Hamiltonians $h_k$ contain the rates of the elementary local
transitions 
and $b_1$, $b_{L-1}$ account for events at the left respectively 
right boundary. The operators $h_m$ include only operators acting on the
sites $m$, $m+1$ and $m+2$; the operators $b_m$ include only
operators acting on $m$ and $m+1$.
They are formulated using the particle number operator ($n_k$), 
vacancy number operator ($v_k=\1-n_k$),
the particle creation ($s^-_k$) and annihilation ($s^+_k$) operators.
The lower index represents the lattice
site on which the respective operator acts. The diagonal entries of the Hamiltonian have to 
be chosen such that conservation of probability is fulfilled which can be easily constructed
by considering
\begin{equation}
\bra{s}s^-_k=\bra{s}v_k; \qquad \bra{s}s^+_k =\bra{s}n_k.
\end{equation}
For example diffusion to the right ($A\varnothing \to \varnothing A$) 
with rate $D_r$ is written as $h_k=D_r \left(s^+_k s^-_{k+1} - n_k v_{k+1}\right)$.

\section{Product-- and shock--measures}
In what follows $\ket{\cdot}_1$ denotes a probability vector for a single
site and an operator without index a single site operator.
If no correlations are present the probability vectors are simply given by
\begin{equation}
\ket{P(t)}=\ket{\rho}\equiv\ket{\rho}_1\otimes \cdots \otimes \ket{\rho}_1
\end{equation}
with
\begin{equation}
\ket{\rho}_1=\rho \ket{\downarrow}_1 + (1-\rho) \ket{\uparrow}_1=
\left(\begin{array}{c} 1-\rho \\ \rho \end{array}\right).
\end{equation}
In this case the calculation of the 
action of the stochastic
generator $H$ can be simplified by 
using the following identities:
\begin{equation}
s^- \ket{\rho}_1=\frac{1-\rho}{\rho} n \ket{\rho}_1; \qquad s^+ \ket{\rho}_1=\frac{\rho}{1-\rho} v \ket{\rho}_1.
\label{3PWW:eq_diagtrans}
\end{equation}
A product measure $\ket{\rho}$ is a stationary state of the system, if
\begin{equation}
H\ket{\rho}=0.
\label{3PWW:eq_prodmeas}
\end{equation} 

\begin{figure}
\centerline{\epsfig{file=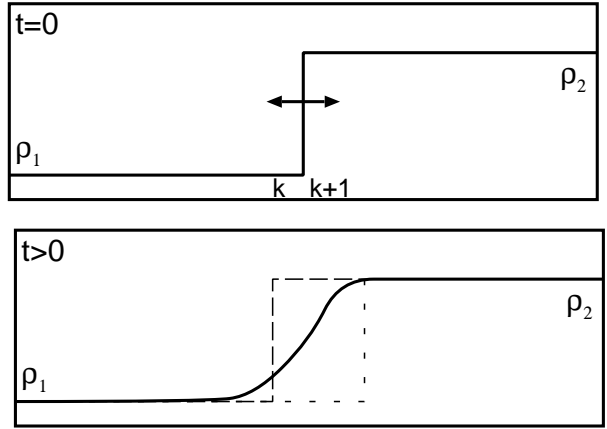,width=8truecm}}
\caption{Ensemble average of a diffusive shock measure (schematically): Due to superposition the 
density profile smears out although the form of the shock is preserved in each 
realization.  
}\label{3PWW:Fig1}
\end{figure}

As a shock measure $\ket{\rho_1,\rho_2,k}$ we define the state which 
is a product measure with density $\rho_1$
up to and including site $k$ and beginning from site $k+1$ a product 
measure with density $\rho_2$:
\begin{equation}
\ket{\rho_1,\rho_2,k}= \ket{\rho_1}_1^{\otimes k} \otimes \ket{\rho_2}_1^{\otimes L-k}.
\label{3PWW:shockmeasure}
\end{equation}
We are interested in those systems, for which the time evolution of 
the shock measure is given by a diffusion
equation (see figure~\ref{3PWW:Fig1})
\begin{eqnarray}
\frac{d}{dt}\ket{\rho_1,\rho_2,k}=- H\ket{\rho_1,\rho_2,k}=& &\delta_1 \ket{\rho_1,\rho_2,k-2}+\delta_2 \ket{\rho_1,\rho_2,k-1}\nonumber\\
       &        +      &  \delta_3 \ket{\rho_1,\rho_2,k+1}+\delta_4 \ket{\rho_1,\rho_2,k+2}\nonumber\\
       &        -      & \delta_5 \ket{\rho_1,\rho_2,k}.
\label{3PWW:shockdiff}
\end{eqnarray}
Starting from an initial density step the system will evolve into
an exact diffusive shock measure defined by
\begin{equation}
\ket{P(t)}=\sum_l p_l(t) \ket{\rho_1,\rho_2,l}
\end{equation} 
which is a time dependent superposition of sharp shocks weighted with $p_l(t)$.
This can be seen by solving
Eq.~(\ref{3PWW:shockdiff}) for an initial shock at position $k$:
\begin{equation}
\begin{split}
\ket{\rho_1,\rho_2,k,t} &= \sum_l G(l,t\vert k,0) \ket{\rho_1,\rho_2,l}\\
G(l,t\vert k,0)&=\frac{1}{2\pi} \int_{-\pi}^{\pi} dp \exp(-\epsilon_p t+i (k-l)p)\\
\epsilon_p&=-(\delta_1 \exp(-2ip)+\delta_2 \exp(-ip)+\delta_3 \exp(ip)+\delta_4 \exp(2ip)
-\delta_5). 
\label{3PWW:diff_sol}
\end{split}
\end{equation}
$G(l,t\vert k,0)$ is the Green's function for this problem, for large arguments $(k-l)$ and
late times it approaches a Gaussian, as expected for a diffusion problem. From 
Eq.~(\ref{3PWW:diff_sol}) we read off $p_l(t)=G(l,t\vert k,0)$.

If we choose a shock measure (\ref{3PWW:shockmeasure}) as initial condition of a system 
which obeys Eq.~(\ref{3PWW:shockdiff}) the form of the shock is preserved in
time but due to the diffusion the state of the system will evolve into a superposition
of shocks. Thus when performing an ensemble average the density profile is not a
sharp step but smears out in time (see figure \ref{3PWW:Fig2}) as seen 
in Monte--Carlo (MC) simulations. Nevertheless a typical configuration of
a single systems shows a sharp shock.

Eq.~(\ref{3PWW:shockdiff}) directly gives the diffusion coefficient $D_s$
and the shock velocity $v_s$:
\begin{equation}
\begin{split}
D_s&=2\delta_1+\delta_2+\delta_3+2\delta_4\\
v_s&=\delta_3+2\delta_4-2\delta_1-\delta_2.
\end{split}
\end{equation} 

The product measures to the left and to the right of the shock position are only
possible for a special choice of boundary dynamics $b_1$ and $b_{L-1}$. A possible
choice is always those boundary dynamics which is the effect of reservoirs with
densities $\rho_1$ respectively $\rho_2$ \cite{Anta00}: One imagines that the lattice is
extended by two additional sites at each boundary and one calculates the action
of $h_{-1}$ and $h_{-2}$ ($h_{L-1}$ and $h_{L}$) assuming that the sites are 
occupied according to the product measures in the bulk and determines the effective
rates for the boundary action. Thus, $n_{-1}$ and $s^{+}_{-1}$ 
are substituted by $\rho_1$, $v_{-1}$ and $s^{-}_{-1}$ by $(1-\rho)$ -- the other
sites are handled accordingly.
In this paper we always consider the limit $L\to\infty$ such that the shock is assumed
not to hit the boundaries.  

Given specific microscopic processes we test for the existence of diffusive shock 
measures by the following
procedure:
\begin{enumerate}[a)]
\item Check whether $\ket{\rho_1}$ is a product measure of the system with periodic boundaries. 
\label{3PWW:enum_a}
\item Set up the boundary processes $b_1$ such that 
\begin{equation}
\left( \sum_{m=1}^k h_m+b_1 \right) \ket{\rho_1,\rho_2,k}= c_{k-1}\ket{\rho_1,\rho_2,k}
\end{equation}
where $c_{k-1}$ acts on sites $k-1$ to $k+2$.
\label{3PWW:enum_b}
\item Check whether $\ket{\rho_2}$ is a product measure of the system with periodic boundaries.
\label{3PWW:enum_c}
\item Set up the boundary processes $b_{L-1}$ such that 
\begin{equation}
\left( \sum_{m=k+1}^{L-2} h_m+b_{L-1} \right) \ket{\rho_1,\rho_2,k}= d_{k+1}\ket{\rho_1,\rho_2,k}
\end{equation}
where $d_{k+1}$ acts on sites $k+1$ to $k+2$.
\label{3PWW:enum_d}
\item Check whether 
\begin{eqnarray}
\left( c_{k-1}+d_{k+1} \right)\ket{\rho_1,\rho_2,k}=&&\delta_1 \ket{\rho_1,\rho_2,k-2}+\delta_2 \ket{\rho_1,\rho_2,k-1}\nonumber\\
       &        +      &  \delta_3 \ket{\rho_1,\rho_2,k+1}+\delta_4 \ket{\rho_1,\rho_2,k+2}\nonumber \\
       &        -      & \delta_5 \ket{\rho_1,\rho_2,k}
\label{3PWW:eq_e}
\end{eqnarray}
\label{3PWW:enum_e}
\end{enumerate}
In detail we do the following: After having set up the local Hamiltonians $h_k$ the action
on a product measure is brought into a diagonal form as described above,
\begin{equation}
h_k \ket{\rho}=h^\text{diag}_k\ket{\rho}
\end{equation}
where $h^\text{diag}_k$ contains only the operators $n_m$ and $\1$ ($v$ can be eliminated 
using $v=\1-n$) acting on 
sites $m=\{k,k+1,k+2\}$. 

The requirement of a product measure (\ref{3PWW:enum_a}) leads to the condition that
the application of the Hamiltonian of the periodic system has to produce 
``telescope''--sums of diagonal operators which results in five equations of the
rates for each density.

Condition \ref{3PWW:enum_e}) can be checked using the identity
\begin{equation}
\ket{\rho_1,\rho_2,k-1}=\left(\frac{1-\rho_2}{1-\rho_1} v_k + \frac{\rho_2}{\rho_1} n_k \right)
                        \ket{\rho_1,\rho_2,k}
\label{3PWW:eq_km}
\end{equation}
and analogous identities for $\ket{\rho_1,\rho_2,k-2}$, $\ket{\rho_1,\rho_2,k+1}$ and
$\ket{\rho_1,\rho_2,k+2}$. A comparison of coefficients of the (only diagonal) operators
on the left and right hand site of Eq.~(\ref{3PWW:eq_e}) then leads to another
seven equations of the rates, $\rho_i$ ($i=1,2$) and $\delta_j$ ($j=1,\ldots,5$).
Conservation of probability implies 
\begin{equation}
\delta_5=\sum_{i=1}^{4}\delta_i,
\end{equation}
which is not an additional equation but simplifies the calculation.

If one of the densities is either zero or one, the Hamiltonian cannot be brought to
diagonal form using Eqs.~(\ref{3PWW:eq_diagtrans}) and (\ref{3PWW:eq_km}). But in this
case the action of the Hamiltonian simplifies and an analogous comparison of 
coefficients of creation and annihilation operators is possible.   

\section{Classification of models}
In order to find the models with three--site interactions which exhibit shock diffusion
one could in principle try to solve Eq.~(\ref{3PWW:shockdiff}) for the 
general Hamiltonian. But as there are 56 microscopic processes (transitions
from any of the $2^3=8$ states to any different state) this task is tedious.
  
We facilitate the procedure by writing a computer program which does all the
symbolical calculation (transformation of the Hamiltonian into diagonal form, gathering
of coefficients) and sets up the constituting set of equations. These
equations are then solved by standard mathematical software. Still, the
general solution of the problem is far too complex to extract useful informations.
It is hence useful to investigate physically motivated 
sub--classes, as done in the following. 

\subsection{Shocks from $\rho_1=0$ to $\rho_2=1$}
The completely empty lattice, $\rho=0$, and the fully occupied one, $\rho=1$,
are nonfluctuating states. They are the two ground states of the zero--temperature
Ising model.
In this case only few processes play a role. All processes starting from
$\0\0\0$ or $AAA$ are forbidden because otherwise product measures with
densities $0$ and $1$ would not be stationary solutions of the system. The processes
which act on the configurations $\0 A\0$, $A\0\0$, $A\0 A$ and $AA\0$ are
ineffectual because none of these configurations is possible, if the system
is initialized with a shock measure. The remaining processes are
\begin{equation}
\begin{split}
\0\0 A \overset{K_0}{\to} A A A; \qquad
\0\0 A \overset{K_1}{\to}\0 A A; \qquad 
\0\0 A \overset{K_2}{\to}\0\0\0;  \\
\0 A A \overset{K_3}{\to} A A A; \qquad 
\0 A A \overset{K_4}{\to}\0\0 A; \qquad 
\0 A A \overset{K_5}{\to}\0\0\0.
\end{split}
\end{equation}
The diffusion constants are then given by
\begin{equation}
\delta_1=K_0; \qquad \delta_2=K_1+K_3; \qquad \delta_3=K_2+K_4; \qquad \delta_4=K_5.
\end{equation}
The solution for a shock between $\rho_1=1$ and $\rho_2=0$ is obtained by exchanging
the roles of particles and holes ($A\leftrightarrow\0$).

This dynamics is a generalized zero temperature Ising model where the domain wall
between spin up and spin down regions diffuses freely. 
As the space symmetric processes have no influence on the upward shock
they can be included giving rise to a model allowing for both, upward
and downward shocks. All ten rates are independent such that the drift of
the domain walls can be chosen freely.

\subsection{Shocks from $\rho_1\in(0;1)$ to $\rho_2\in (0;1)$}
\label{3PWW:sec4b}
For each Hamiltonian with three--site interactions there is either none, exactly one 
product measure with a density in the open interval $\rho\in(0;1)$, or infinitely many. 
The latter is only possible for particle--conserving Hamiltonians. 
The proof of this assertion is possible for the case of three--site interactions, but
rather technical and is therefore omitted here.
It is based on analyzing the conditions on the rates and densities
for the existence of a product measure Eq.~(\ref{3PWW:eq_prodmeas})
for the general
three--site interaction Hamiltonian.

Instead we present a general consideration
why the existence of two fluctuating stationary product measures is impossible for non particle--conserving
Hamiltonians. This is not intended to be a mathematical rigorous proof, it is rather
a physical consistency check. This consideration is related to the positive
rates conjecture, see Refs.~\cite{Gray01,Gacs01}.

Let us assume that there are two fluctuating stationary densities $\rho_1\ne\rho_2$ for 
a non particle--conserving Hamiltonian.
Then due to fluctuations there is a finite probability that in the stationary state with
density $\rho_1$ a region of density $\rho_2$ emerges. As $\rho_1$ is a stationary density this
region has to be suppressed and vanishes again. Consequently the domain walls $\rho_1\vert\rho_2$
and $\rho_2\vert\rho_1$ are moving toward each other shortening the domain of $\rho_2$. 
But if this is the case $\rho_2$ cannot
be a stationary state, as a region of $\rho_1$ emerging in a phase of $\rho_2$ is growing. 
Hence the assumption of the existence of two fluctuating stationary densities was wrong.

Note that Gacs error correcting model \cite{Gray01,Gacs01} does not represent
a counter example for this statement. It is crucial that product measures
are considered: For this case no dynamics is able to determine the length of
a one--dimensional domain -- the domain growth is independent of the domain
size, while phase separation requires a faster growth of larger domains.

\begin{figure}[tn]
\vspace{5mm}
\centerline{\epsfig{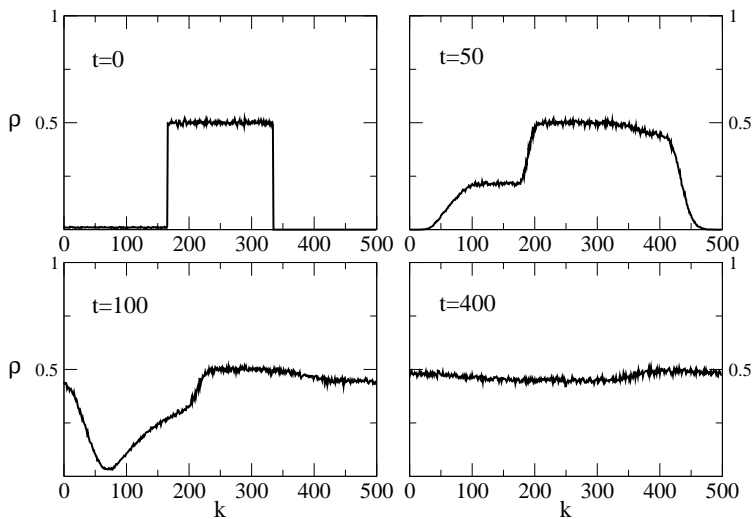}}
\caption{Monte Carlo simulation of the model $A\0\0\to AAA; \,\, AAA\to \0\0A$, the number of sites
is 500 and an average over 5000 systems was performed, the boundary conditions are periodic. 
As initial condition 
we took a shock measure with steps at site 166 from $0.01$ to $1/2$ and
at site 333 from $1/2$ to $0$. The left step is not stable because the product measure
with $\rho_1=0.01$ is not a stationary state and the right step is not stable because
there the diffusion equation is not fulfilled.
}\label{3PWW:Fig2}
\end{figure}

In Fig.~\ref{3PWW:Fig2} the time evolution of a non particle--conserving system is shown.
The particle density 
in the low density region
is not stationary and increases until the first shock
vanishes. The second shock disappears due to another mechanism, here the diffusion equation of
the shock is not fulfilled and consequently the shock dissolves.

The situation is basically different if particle--conserving Hamiltonians are considered.
In this case the system is not ergodic and hence a region of different density cannot
evolve in the bulk of the system. Consequently several stationary densities are possible. 

As non--conserving Hamiltonians do not allow for the existence of 
two fluctuating phases we only need to investigate particle--conserving Hamiltonians and recover
the solution of the investigation of nearest neighbor interactions,
the asymmetric simple exclusion process (ASEP) \cite{Kreb03}:
\begin{equation}
\begin{split}
 A\0 \overset{D_R}{\to}\0 A; &\qquad \0 A \overset{D_L}{\to}A \0\\
D_R \frac{\rho_1}{1-\rho_1}=&D_L \frac{\rho_2}{1-\rho_2}\\
\delta_1=0;\qquad\delta_2=\frac{1-\rho_1}{1-\rho_2} D_L; &\qquad
\delta_2=\frac{1-\rho_2}{1-\rho_1} D_R; \qquad \delta_4=0
\end{split}
\end{equation}

The inclusion of next--nearest neighbor interaction does not lead to further models in this case.

\subsection{Shocks from $\rho_1=0$ to $\rho_2\in(0;1)$}
While the existence of two fluctuating stationary states is impossible, the existence of
one fluctuating and one non fluctuating phase is easy to construct. The density $\rho_1=0$
is stationary if all processes from the empty lattice, $\0\0\0$, are prohibited. This is a violation of the assumption underlying the positive rates conjecture as certain types of
fluctuations are absent. Therefore two stationary states
are possible, however, the nonfluctuating phase is unstable and the
fluctuating phase will always enlarge at the expense of the nonfluctuating one.
 
For the existence of a
product measure $\rho_2\in(0;1)$ it is then necessary that as well no process
to the empty lattice is present. Apart from these constraints no further processes
can be excluded a priori. We restrict ourselves to the case of a shock between
$\rho_1=0$ to $\rho_2\in(0;1)$. The cases of a shock from $\rho_1=1$ to $\rho_2\in(0;1)$
can be obtained by exchanging particles and holes, and the case of the fluctuating
phase to the left can be obtained by a parity transformation.

As argued above, the degrees of freedom are too many for a complete investigation and we
classify the systems by the symmetries charge ($C$), parity ($P$) and time ($T$).
In this context, charge symmetry is the invariance of the microscopic processes under the
exchange of particles and holes ($A\leftrightarrow\0$), i.e., for each process in the model
there exists the $C$-symmetric one with the same rate.
Although this picture is quite artificial when applied to particles it is natural
in the language of spins where in the absence of an external field the symmetry between
the up and down spin is obvious.
 Parity symmetry is the invariance
of the microscopic processes under the exchange of left and right, i.e., for each process
in the model there exists the $P$-symmetric one with the same rate. Time symmetry is the
invariance under the transformation of $t\to -t$. A stochastic
model is $T$--invariant if detailed balance with respect to its stationary distribution is 
fulfilled \cite{Kamp01}, i.e., these models are able to reach an equilibrium steady state. 
This is the case if for all states $\eta_1$,$\eta_2$ the transition rates 
$w_{\eta_1\to\eta_2}$, $w_{\eta_2\to\eta_1}$ and 
probabilities of finding the system in the configurations $P(\eta_1)$, $P(\eta_2)$, obey
the equation $w_{\eta_1\to\eta_2}P(\eta_1)=w_{\eta_2\to\eta_1}P(\eta_2)$, 
i.e., there is no net current between states.
If the stationary state of a system is a product measure, the validity of detailed 
balance is easily checked or refuted, because the calculation of the probabilities
of the system to be in a specific states is trivial and additionally the 
absence of correlations permits to investigate only the local rates instead of the 
configurations of the whole lattice.

One has to distinguish between a combined symmetry, for example
$PT$, from an independent symmetry, here expressed by the symbol
'$\wedge$', for example $P\wedge T$. While in the former case the system
is invariant after applying the symmetries one after each other, in the
latter case the system is invariant under the symmetries applied
each by themselves.

\subsubsection{$C\wedge P\wedge T$ symmetric systems with shock measures}
The 56 possible three--site interaction transitions can be arranged in 11 minimal
models which obey $C\wedge P \wedge T$ symmetry each (but are not necessarily described
by exact diffusive shock measures). 
If we exclude those which 
have transitions involving the configurations $\0\0\0$ or $AAA$ only 6 remain. 
There are 63 combinations which can be build out of 6 elements.
We checked all combinations and found that there is only one
model which is $C\wedge P\wedge T$ symmetric and has an exact diffusive shock
measure as solution:
\begin{equation}
\begin{split}
A\0 \overset{1}{\to}\0 A; &\qquad \0 A \overset{1}{\to}A \0;\\ 
\0\0 A \overset{1}{\to}\0 A A; \qquad \0 A A \overset{1}{\to}\0\0 A; &\qquad
A\0 \0 \overset{1}{\to}A A \0; \qquad  A A\0 \overset{1}{\to} A\0\0;\\
\rho_2=1/2; \qquad \delta_1=\delta_4=0;&\qquad \delta_2=2; \qquad \delta_3=1.
\end{split}
\end{equation}
In this model diffusion is combined with branching processes which are 
only possible to a neighboring lattice site if the subsequent site
is empty and its reversal, the coalescence process.

\begin{figure}[tn]
\vspace{5mm}
\centerline{\epsfig{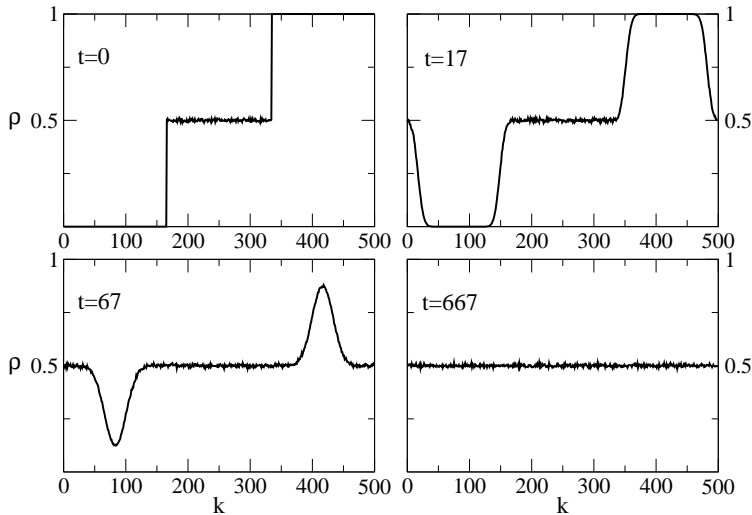}}
\caption{Monte Carlo simulation of the $C\wedge P\wedge T$ symmetric model, the number of sites
is 500 and an average over 10000 systems was performed. As initial condition 
we took a shock measure with steps at site 166 from $0$ to $1/2$ and
at site 333 from $1/2$ to $1$.
}\label{3PWW:Fig3}
\end{figure}

Due to the $C\wedge P$ symmetry this model has the property that both the shock
from $0$ or $1$ to $1/2$ and the shock from $1/2$ to $0$ or $1$ is stable. Hence double
shocks $0\vert \frac{1}{2} \vert 0$ and $0\vert \frac{1}{2} \vert 1$ are possible.
We calculated the time evolution
of an initial double shock ($0\vert\frac{1}{2}\vert 1$) in a periodic system by a Monte 
Carlo simulation as shown
in Fig.~\ref{3PWW:Fig3}. 
As calculated the shock moves with a drift of $1$ and such that always
the fluctuating phase penetrates the non--fluctuating phase. 
The shock $1\vert 0$ between site $500$ and $0$ evolves into two steps,
$1\vert \frac{1}{2}$ and $\frac{1}{2}\vert 0$, since these shocks drift into
the nonfluctuating phase and split up. As a consequence of the periodic boundary conditions
the two shock fronts moving into the non--fluctuating phase coalesce after finite time and 
the non--fluctuating phase vanishes. This is a consequence of the
stability of the fluctuating phase as argued in section \ref{3PWW:sec4b}.
A detailed discussion of the reaction of shock fronts will be given
in \ref{3PWW:secreact}.

\subsubsection{$C\wedge P$ symmetric systems with shock measures}
There are 17 minimal models obeying $C\wedge P$ symmetry, 
if we exclude those which have transitions involving the configurations
 $\0\0\0$ and $AAA$ only 8 remain.
Thus in this case 255 combinations have to be checked, but no additional model besides
the $C\wedge P\wedge T$ model can be found.  

\subsubsection{$P\wedge T$ symmetric systems}
There are 18 minimal models obeying $P\wedge T$ symmetry, 
if we exclude those which have transitions involving the configurations $\0\0\0$ only 13 remain. Thus in this
case 8191 combinations have to be checked. We found 14 models, as presented in 
the appendix. Due to the $P$ symmetry in each of these models
downward shocks are also stable. 

In Fig.~\ref{3PWW:Fig4} we show a Monte Carlo simulation of the model D (see
appendix) with $\omega=1/2$ on a ring. As predicted two aspects can be observed:
First, both the upward and the downward shock are stable. Second, the fluctuating phase 
spreads into the nonfluctuating one until the latter finally vanishes. We remind that due
to the superposition of shocks the ensemble average does not exhibit a sharp step although
each single realization does.

\begin{figure}[tn]
\vspace{5mm}
\centerline{\epsfig{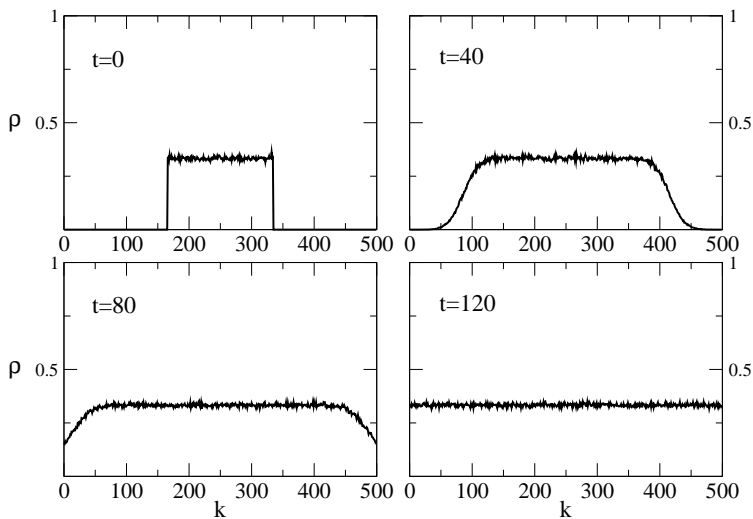}}
\caption{Monte Carlo simulation of the $P\wedge T$ symmetric model D with $\omega=1/2$, the number of sites
is 500 and an average over 5000 systems was performed. As initial condition 
we took a shock measure with steps at site 166 from $0$ to $1/3$ and
at site 333 from $1/3$ to $0$.
}\label{3PWW:Fig4}
\end{figure}

\subsubsection{$C\wedge T$ symmetric systems}
There are 16 minimal models obeying $C\wedge T$ symmetry, 
if we exclude those which include the configurations $\0\0\0$ or $AAA$ only 9 remain.
Thus in this case 511 combinations have to be checked.
Besides the $C\wedge P\wedge T$ model we find the following two models:
\begin{equation}
\begin{split}
\0 A\0 \overset{1}{\to} \0 A A; \qquad \0 A A \overset{1}{\to} \0 A\0; &\qquad
 A\0\0 \overset{1}{\to}  A\0 A; \qquad  A\0 A \overset{1}{\to}  A \0\0; \\
\rho_2=1/2; &\qquad \delta_1=\delta_2=\delta_3=\delta_4=0.
\end{split}
\end{equation}
In this model only branching to the right is possible in the presence of another zero at 
the nearest neighboring site and the corresponding coalescence processes. The domain
wall $0\vert \frac{1}{2}$ is not fluctuating even though the domain is.

The second model is a combination of branching and coalescence processes to both directions
\begin{equation}
\begin{split}
\0\0 A \overset{1}{\to} \0 A A; \qquad \0 A A \overset{1}{\to} \0\0 A; &\qquad
 A\0\0 \overset{1}{\to}  A A\0; \qquad  A A \0 \overset{1}{\to}  A\0\0; \\
 A \0\0 \overset{1}{\to} \0 A A; \qquad \0 A A \overset{1}{\to}  A\0\0; &\qquad
\0 A\0 \overset{1}{\to}  A\0 A; \qquad  A \0 A \overset{1}{\to} \0 A\0; \\
\rho_2=1/2; &\qquad \delta_1=\delta_4=0; \qquad \delta_2=\delta_3=2.
\end{split}
\end{equation}
The domain wall performs an unbiased diffusion with diffusion constant $D=2$.

For these two $C\wedge T$ models a downward shock has not necessarily to be stable as
it is the case for the models which are $P$ symmetric. Indeed, a downward shock
is {\em not} stable in the two models because otherwise the space reflected versions
of the processes would show a stable upward shock and would constitute additional
$C\wedge T$ models.

\subsubsection{Some further models}
Here we present some models which do not belong to the classes presented above. 

The following model is the totally asymmetric exclusion process combined
with activated Langmuir kinetics used in Ref.~\cite{Rako03} to show hysteresis in 
driven diffusive systems:
\begin{equation}
\begin{split}
 A\0 \overset{1}{\to} \0 A;&\\
A \0 A \overset{\omega_a}{\to} A A A; &\qquad A A A \overset{\omega_d}{\to} A\0 A;\\
\rho_2=\frac{\omega_a}{\omega_a+\omega_d}; &\qquad \delta_1=\delta_2=\delta_4=0; \qquad
\delta_3=(1-\rho_2).
\end{split}
\end{equation}

The simplest model with a fluctuating shock front not included in nearest neighbor
interaction models is 
\begin{equation}
A\0\0\to \0\0 A;\qquad \delta_4=(1-\rho_2)^2,
\end{equation}
where particles are only allowed to hop over a vacancy to the right. This model is
$PT$ symmetric, because reversing the direction of the process is equal to exchanging
left and right.

Another simple model is 
\begin{equation}
\begin{split}
A\0\0 \overset{1}{\to} A A A; &\qquad A A A \overset{\omega}{\to} \0\0 A; \\
\rho_2=\frac{\sqrt{\omega}-1}{\omega-1}; &\qquad \delta_1=\delta_2=\delta_3=0; \qquad 
\delta_4=\omega \rho_2^2.
\end{split}
\end{equation}
This model is again $PT$ symmetric by the same argument as in the model presented above.
In this case one has to pay attention to the rates when reversing the time direction, the
time reversed process is:
\begin{equation}
A A A \overset{\omega}{\to} A\0\0; \qquad \0\0 A \overset{1}{\to} A A A.
\end{equation}
For $\omega=1$ the density is $\rho_2=\frac{1}{2}$.

\section{Reactions of domain walls}
\label{3PWW:secreact}
Up to now we have only investigated whether the shock fronts are stable and if so
how they move. The movement of the shocks can be used to describe the dynamics
of the systems as their positions are sufficient to characterize the state of the 
system. This has been used for example in Ref.~\cite{Rako03} where the
dynamics with many degrees of freedom could be reduced to an effective one
particle system. In order to describe a system completely by the position 
of the domain walls we additionally have to investigate how the domain walls
affect each other. 

A special case would be if the domain walls do not interact at all. This means
that their rates do not change in the presence of another wall -- certainly
the possibility of mutual annihilation has to be included as a phase may vanish if
two boundaries meet. 
In this case one could interpret the dynamics as the motion of annihilating random
walkers. For those systems a direct link to free fermion system has been discussed
in Ref.~\cite{Schu01} and
thus, for the unaffected movement of domain walls a description by free fermions
could be possible. To this end one may apply the Jordan--Wigner \cite{Jord28}
transformation which
converts spin--1/2 operators into fermionic creation and annihilation operators.
 
The significance of the Jordan--Wigner transformation in this context is that some 
Hamiltonians of spin--1/2 systems
transform into fermion Hamiltonians which include only bilinear 
expressions of the fermionic operators  -- 
this can be regarded as a system of free fermions. For these Hamiltonians additional techniques
for calculating the dynamical properties are available. 
Interestingly the dynamics of all particle models showing stable shock 
fronts with two--site interactions can be represented by the free motion of domain
walls without interactions \cite{Kreb03}. For the BCRW and the AKGP this is directly related
to the free fermion character of these systems.

It is the purpose of this section to investigate potential relations of the three--site interactions models
found above to fermion systems,
since a link between the domain wall motion and free fermion behavior would be interesting.
We first discuss the dynamics of domain walls in detail and then turn to the transformation
into fermion systems.

\subsubsection{Dynamics of domain walls}
When investigating the interaction of domain walls,
the situation simplifies again by the fact that within the domains the probabilities
are given by a product measure. By this two domain walls may only influence each
other if the distance is smaller than three lattice sites and thus only operators acting
on a small range have to be included.

The first model to be considered is the $C\wedge P\wedge T$ model. In Fig.~\ref{3PWW:Fig3}
on the one hand it is shown how a domain wall $1\vert 0$ splits into two domain walls
$1\vert\frac{1}{2}$ and $\frac{1}{2}\vert 0$ and on the other hand how two domain walls
$\frac{1}{2}\vert 0$ and $0 \vert\frac{1}{2}$ coalesce. These two cases are now studied
in detail analytically.

If the system is characterized simply by the position of the domain walls without
further correlations it will be sufficient to describe the dynamics by states
\begin{equation}
\ket{k,l;\rho_1,\rho_2,\rho_3}=\cdots\otimes\ket{\rho_1}\otimes\cdots\underset{k}{\ket{\rho_1}}
                               \otimes\underset{k+1}{\ket{\rho_2}}\otimes\cdots
                               \otimes\underset{l-1}{\ket{\rho_2}}
                               \otimes\underset{l}{\ket{\rho_3}}\otimes\cdots.
\end{equation}
In order to describe the evolution of the
step $1\vert 0$ we define
\begin{equation}
\ket{k,l}\equiv\ket{k,l;1,\frac{1}{2},0}. 
\end{equation} 
In the following the densities are omitted for the sake of simplicity. By applying the
Hamiltonian of the $C\wedge P\wedge T$ model on this state one gets
\begin{equation}
\frac{\partial}{\partial t} \ket{k,l}=\left\{\begin{array}{cc}
                                         4 \ket{k-1,l+1}- 4\ket{k,l} & l-k=1 \\[0.1cm]
                                         2 \ket{k-1,l}+2 \ket{k,l+1}- 4\ket{k,l} & l-k=2 \\[0.1cm]
          2\ket{k-1,l}+\ket{k+1,l}+2\ket{k,l+1}+\ket{k,l-1}-\lefteqn{6\ket{k,l}} & \\ 
                                                                          &   l-k\ge 3.
\end{array}
\right.
\end{equation}
\begin{figure}[tn]
\centerline{\epsfig{file=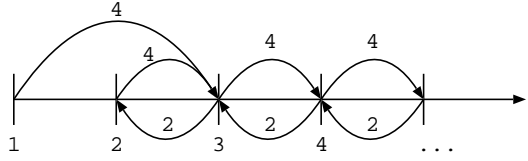,width=7truecm}}
\caption{The difference $l-k$ of the two shock fronts in the state $\ket{k,l}$ perform a biased
random walk.
}\label{3PWW:fig:rw}
\end{figure}

Thus the time evolution of the system can be completely described by the movement of the domain walls, no
additional correlations evolve.
Starting from $1\vert 0$ both domain walls simultaneously move by one lattice site creating a 
fluctuating domain of two sites with density $\frac{1}{2}$. The time evolution of a state in which the domain walls
are separated by a lattice site is given by the separate movement of both domain walls by one lattice site. 
A state in which the domain walls are separated by two or more lattice sites evolves simply by 
the rates $\delta_2=2$ and $\delta_3=1$ as calculated before. The time evolution of the distance of the
domain walls $l-k$ is illustrated in Fig.~\ref{3PWW:fig:rw}. Once the domain walls are separated it
is not possible that they coalesce again as the $l-k=1$ is an isolated point. Thus the movements of the
domain walls are not independent, a repulsive interaction is present.

Next, the interaction of the shock fronts $\frac{1}{2}\vert 0$ and $0\vert\frac{1}{2}$ shall be investigated. To
this end we now define
\begin{equation}
\ket{k,l}\equiv\ket{k,l;\frac{1}{2},0,\frac{1}{2}},
\end{equation}
and test again whether the dynamics can be described in terms of the $\ket{k,l}$. Applying the Hamiltonian
on the state $\ket{k-1,k+1}$ shows that additional correlations appear and that consequently the time
evolution cannot be described simply by the location of the domain walls. 

Nevertheless it is instructive to analyze the appearance of the correlations in detail in
order to reveal the link to free fermion systems.
The correlations can be compensated by including the process 
$A\leer A \leftrightharpoons AAA$.
By this the $C$ symmetry is broken (the $\rho=1$ phase is not stable anymore), but it is
still a $P\wedge T$ model.

 Choosing the rate for the forward and backward reaction to be $2$ 
one gets:
\begin{equation}
\frac{\partial}{\partial t} \ket{k,l}=\left\{\begin{array}{cc}
                                         0 & l-k=1 \\[0.1cm]
                                         4 \ket{\frac{1}{2}}+ \ket{k-1,l}+\ket{k,l+1}- 6\ket{k,l} & l-k=2 \\[0.1cm]
          \ket{k-1,l}+\ket{k,l+1}+2\ket{k+1,l}+2\ket{k,l-1}-\lefteqn{6\ket{k,l}} & \\
                                                                           &   l-k\ge 3,
\end{array}
\right.
\end{equation}
where $\ket{\frac{1}{2}}=\ket{k,l}|_{k-l=1}$ is the product measure with density $\rho=\frac{1}{2}$. This state is stationary which is recovered by the vanishing time derivative
for $k-l=1$.
The time evolution
of the system can be described completely by the movement of the shocks and their dynamics
is independent of each other until they meet, then both are annihilated. 

Consequently this 
system is a candidate for the description by free fer\-mi\-ons. However, by including the process 
$A\leer A \leftrightharpoons AAA$ we get the model~\ref{3PWW:modelBCRW} of the appendix which
is the BCRW -- only a two--site interaction model. 
The transformation of the BCRW into a free fermion system is known 
\cite{Schu01,Kreb03}.  \\

As a second example the model~\ref{3PWW:modelJ} of the appendix is chosen as 
it is one of the simplest models. Again we define
\begin{equation}
\ket{k,l}\equiv\ket{k,l;\rho,0,\rho},
\end{equation}
and apply the Hamiltonian. The description by the states $\ket{k,l}$ is only closed if we set
the parameter of the model $w=1$ for which $\rho=1/2$ and some of the rates vanish. One gets
\begin{equation}
\frac{\partial}{\partial t} \ket{k,l}=\left\{\begin{array}{cc}
                                         0 & l-k=1 \\[0.1cm]
                                         4 \ket{\frac{1}{2}}+ \ket{k-1,l}+\ket{k,l+1}- 6\ket{k,l} & l-k=2 \\[0.1cm]
          \ket{k-1,l}+\ket{k,l+1}+2\ket{k+1,l}+2\ket{k,l-1}-\lefteqn{6\ket{k,l}} & \\
                                                                           &   l-k\ge 3,
\end{array}
\right.
\end{equation}

Next the time evolution of a state
\begin{equation}
\ket{k,l}\equiv\ket{k,l;0,\frac{1}{2},0}
\end{equation}
is investigated. 

We find
\begin{equation}
\frac{\partial}{\partial t} \ket{k,l}=\left\{\begin{array}{cc}                                          0 & l-k=1 \\[0.1cm]
          \ket{k+1,l}+\ket{k,l-1}+2\ket{k-1,l}+2\ket{k,l+1}-\lefteqn{6\ket{k,l}} & \\
                                                                           &   l-k\ge 3.
\end{array}
\right.
\end{equation}
For $k-l=2$ it turns out that the time evolution of this state cannot
be described by a superposition of shock measures, i.e. additional correlations 
emerge. Thus
no independent movement of the shock fronts is possible in this case. 
This suggests that this is not a free fermion model.

\section{Conclusions}
In this paper we have investigated exact diffusive shock measures in one--dimensional 
reaction diffusion systems with
next nearest neighbor interactions and open boundaries.  
We distinguish the following three cases:

1. The connection of two non--fluctuating phases, the two densities are 0 and 1. 
The conditions that both
states are stable exclude many models and we find many next nearest neighbor models
as solution generalizing the Glauber Ising model at zero temperature.

We restricted ourselves to the case of completely ordered connected phases as initial
conditions. It would be interesting to investigate how a system evolves out of random
initial conditions. In this scenario coarsening of the ordered domains or the emergence
of a third stationary state which is fluctuating is possible.
 
2. The connection of two fluctuating phases, both densities are between 0 and 1. 
It is argued that 
in general non--conservative models cannot have two fluctuating product measures as solution,
in agreement with the positive rates conjecture.
Consequently only conservative models have to be investigated and we recover the ASEP as the most
general solution. Hence, the inclusion of three--site interactions does not lead to 
models not known
from the investigation of nearest neighbor interactions.  

3. The connection of a non--fluctuating phase to a fluctuating phase, one density 0 or 1 and one
between 0 and 1. In this case numerous models exist and we classify the systems with respect to 
their symmetry. There is only one model which is $C\wedge P\wedge T$ invariant, and this is as well the only
model which is $C\wedge P$ invariant. Two additional models are found that are $C \wedge T$ invariant and 14 
models are found that are $P\wedge T$ invariant.

We stress that the mechanisms of exact shock measures are {\em not} suitable to construct a (one--species) model which shows phase separation on a ring. On the one hand, although models with one fluctuating phase and one non--fluctuating phase allow for stable up-- and downward--shocks, the non--fluctuating phase will always vanish on a ring because the two shock fronts always enlarge the active region. On the other hand
conservative models, which in principle allow for shocks between two fluctuating phases, are
unable to
show both up-- and downward--shocks. This is a consequence of the collective
velocity $v_c(\rho)$ which
describes the movement of the center of mass of a disturbance in a region of a
certain density $\rho$.
In order that a shock is stable a disturbance has to tend toward the shock,
$v_c(\rho_1)>v_s>v_c(\rho_2)$,
where $v_s$ is the shock velocity. Obviously this equation can only hold either
for the upward-- or
the downward--shock. Note that this consideration only holds for short--ranged,
homogeneous one--species models, it is
known that phase separation is possible in models with defects \cite{Mall96} or
several species of particles \cite{Evan98a,Kafr03}.

Although it is shown that double shocks $0\vert \rho \vert 0$ are possible we argue 
that this cannot
be used to construct models which show phase separation in one--dimensional, short--ranged 
periodic
systems with a single species.

We have also investigated the influence of shock fronts on each other. A case of special
interest is when the fronts move independently, except for the possibility of mutual
annihilation. In the $C\wedge P\wedge T$ model this cannot be observed, it turns out that
it has to be combined with an additional process violating $C$ symmetry.
But by this the BCRW which is a two--site interaction model is recovered
for which the independence of shock fronts is known. 

We conclude that there is no direct connection of models whose time evolution is given
by exact diffusive shock measures and free fermion systems.

\section{Appendix: $P\wedge T$ models}
The 14 $P\wedge T$ models are:
{
\begin{enumerate}[Model A:]
\item 
\begin{equation}
\begin{split}
 A\0 A \overset{\omega_a}{\rightarrow} A A A;& \qquad A A A \overset{\omega_d}{\rightarrow} A\0 A\\
%10
\rho=\frac{\omega_a}{\omega_a+\omega_d}; &\delta_1=\delta_2=\delta_3=\delta_4=0
\end{split}
\end{equation}
The shock position is fixed without fluctuations in this model, but the model is not ergodic.

\item
\begin{equation}
\begin{split}
A\0 \overset{1}{\rightarrow} \0 A; & \qquad \0 A \overset{1}{\rightarrow} A \0;\\
%DIFF
\0\0 A \overset{\omega}{\to}\0 A A;& \qquad \0 A A \overset{1}{\to}\0\0 A;\\
A\0 \0 \overset{\omega}{\to}A A \0;& \qquad  A A\0 \overset{1}{\to} A\0\0;\\
%2
\rho_2=\frac{\omega}{\omega+1}; \qquad \delta_1=\delta_4=0;&\qquad \delta_2=\frac{1}{1-\rho_2}; \qquad \delta_3=1
\end{split}
\end{equation}
In the case $\omega=1$ this model is $C$--invariant and we recover the $C\wedge P\wedge T$ model.

\item
\label{3PWW:modelBCRW}
\begin{equation}
\begin{split}
A\0 \overset{1}{\to} \0 A; & \qquad \0 A \overset{1}{\to} A \0\\
A\0 \overset{\omega}{\to} A A;& \qquad \0 A \overset{\omega}{\to} A A;\\
A A \overset{1}{\to} A \0; &\qquad A A \overset{1}{\to} \0 A;\\
\rho_2=\frac{\omega}{\omega+1}; \qquad \delta_1=\delta_4=0;&\qquad \delta_2=\frac{1}{1-\rho_2};\qquad \delta_3=1;
\end{split}
\end{equation}
This model is a purely two--site interaction model and known as the branching coalescing
random walk.
It can be obtained by combining model A and B.

\item
\begin{equation}
\begin{split}
\0\0 A \overset{1}{\to}\0 A\0; &\qquad A \0\0 \overset{1}{\to}\0 A\0;\\
\0 A\0 \overset{1}{\to}\0\0 A; &\qquad \0 A\0 \overset{1}{\to} A\0\0;\\
\0\0 A \overset{\alpha\omega}{\to}\0 A A; &\qquad \0 A A \overset{\alpha}{\to}\0\0 A; \\
 A\0\0 \overset{\alpha\omega}{\to} A A\0; &\qquad  A A\0 \overset{\alpha}{\to} A\0\0;\\
\0 A\0 \overset{\omega}{\to} A\0 A; &\qquad  A\0 A \overset{1}{\to}\0 A\0; \\
\0 A A \overset{\frac{1}{1-\omega}}{\to} A A\0; &\qquad  A A\0 \overset{\frac{1}{1-\omega}}{\to}\0 A A;\\
\0 A\0 \overset{\frac{\omega^2}{1-\omega}}{\to} A A A; &\qquad  A A A \overset{\frac{1}{1-\omega}}{\to}\0 A\0;\\
\alpha=\frac{(1-2\,\omega)}{1-\omega}; &\qquad \omega\le \frac{1}{2}\\
\rho_2=\frac{\omega}{\omega+1}; \qquad \delta_1=\delta_4=0;&\qquad \delta_2=\frac{2}{1-\rho_2}; \qquad \delta_3=2
\end{split}
\end{equation}
For $\omega=\frac{1}{2}$ we get $\alpha=0$ and some of the rates vanish.

\item
\begin{equation}
\begin{split}
\0\0 A \overset{1}{\to}\0 A\0; &\qquad A \0\0 \overset{1}{\to}\0 A\0;\\
\0 A\0 \overset{1}{\to}\0\0 A; &\qquad \0 A\0 \overset{1}{\to} A\0\0;\\
\0\0 A \overset{\omega}{\to}\0 A A; &\qquad \0 A A \overset{1}{\to}\0\0 A;\\
 A\0\0 \overset{\omega}{\to} A A\0; &\qquad  A A\0 \overset{1}{\to} A\0\0;\\
\0 A\0 \overset{\omega}{\to} A\0 A; &\qquad  A\0 A \overset{1}{\to}\0 A\0; \\
\0 A A \overset{1}{\to} A A\0; &\qquad  A A\0 \overset{1}{\to}\0 A A;\\
\0 A A \overset{\omega}{\to} A A A; &\qquad  A A\0 \overset{\omega}{\to} A A A; \\
 A A A \overset{1}{\to} \0 A A; &\qquad  A A A \overset{1}{\to} A A\0;\\
\rho_2=\frac{\omega}{\omega+1}; \qquad \delta_1=\delta_4=0;&\qquad \delta_2=\frac{2}{1-\rho_2}; \qquad \delta_3=2
\end{split}
\end{equation}

\item
\begin{equation}
\begin{split}
\0\0 A \overset{1}{\to}\0 A\0; &\qquad A \0\0 \overset{1}{\to}\0 A\0;\\
\0 A\0 \overset{1}{\to}\0\0 A; &\qquad \0 A\0 \overset{1}{\to} A\0\0;\\
%1
\0 A\0 \overset{\omega(1-2\omega)}{\to}\0 A A; &\qquad \0 A\0 \overset{\omega(1-2\omega)}{\to} A A\0;\\
 \0 A A \overset{1-2\omega}{\to} \0 A\0; &\qquad  A A\0 \overset{\frac{1-2\omega}{\omega^2}}{\to} \0 A\0;\\
%4
\0 A\0 \overset{\omega}{\to} A\0 A; &\qquad  A\0 A \overset{1}{\to}\0 A\0; \\
%5
\0 A A \overset{2\omega+1}{\to} A A\0; &\qquad  A A\0 \overset{2\omega+1}{\to}\0 A A;\\
%7
\0 A\0 \overset{2\omega^2}{\to} A A A; &\qquad  A A A \overset{2}{\to}\0 A\0; \\
%8
\omega\le \frac{1}{2};&\\
\rho_2=\frac{\omega}{\omega+1}; \qquad \delta_1=\delta_4=0;&\qquad \delta_2=\frac{2}{1-\rho_2}; \qquad \delta_3=2
\end{split}
\end{equation}
For $\omega=\frac{1}{2}$ some rates vanish.

\item
\begin{equation}
\begin{split}
\0\0 A \overset{1}{\to}\0 A\0; &\qquad A \0\0 \overset{1}{\to}\0 A\0; \\
\0 A\0 \overset{1}{\to}\0\0 A; &\qquad \0 A\0 \overset{1}{\to} A\0\0;\\
%1
\0 A\0 \overset{\omega}{\to}\0 A A; &\qquad \0 A\0 \overset{\omega}{\to} A A\0; \\
 \0 A A \overset{1}{\to} \0\0 A; &\qquad  A A\0 \overset{1}{\to} \0 A\0;\\
%4
\0 A\0 \overset{\omega}{\to} A\0 A; &\qquad  A\0 A \overset{1}{\to}\0 A\0; \\
%5
\0 A A \overset{1}{\to} A A\0; &\qquad  A A\0 \overset{1}{\to}\0 A A;\\
%7
\0 A A \overset{2\omega}{\to} A A A; &\qquad  A A\0 \overset{2\omega}{\to} A A A; \\
 A A A \overset{2}{\to} \0 A A; &\qquad  A A A \overset{2}{\to} A A\0;\\
%9
\rho_2=\frac{\omega}{\omega+1}; \qquad \delta_1=\delta_4=0;&\qquad \delta_2=\frac{2}{1-\rho_2}; \qquad \delta_3=2
\end{split}
\end{equation}

\item
\begin{equation}
\begin{split}
\0\0 A \overset{1}{\to}\0 A\0; &\qquad A \0\0 \overset{1}{\to}\0 A\0;\\ 
\0 A\0 \overset{1}{\to}\0\0 A; &\qquad \0 A\0 \overset{1}{\to} A\0\0;\\
%1
\0 A\0 \overset{2\omega(1-\omega)}{\to}\0 A A; &\qquad \0 A\0 \overset{2\omega(1-\omega)}{\to} A A\0;\\
 \0 A A \overset{2(1-\omega)}{\to} \0 A\0; &\qquad  A A\0 \overset{2(1-\omega)}{\to} \0 A\0;\\
%4
\0 A A \overset{1}{\to} A\0 A; &\qquad A\0 A \overset{1}{\to} \0 A A;\\
 A\0 A \overset{1}{\to} A A\0; &\qquad \0 A A \overset{1}{\to} A\0 A;\\
%6
\0 A A \overset{2\omega}{\to} A A\0; &\qquad  A A\0 \overset{2\omega}{\to}\0 A A;\\
%7
\0 A\0 \overset{2\omega^2}{\to} A A A; &\qquad  A A A \overset{2}{\to}\0 A\0; \\
%8
\omega\le 1;&\\
\rho_2=\frac{\omega}{\omega+1}; \qquad \delta_1=\delta_4=0;&\qquad \delta_2=\frac{2}{1-\rho_2}; \qquad \delta_3=2
\end{split}
\end{equation}
For $\omega=1$ some of the rates vanish.

\item
\begin{equation}
\begin{split}
\0\0 A \overset{1}{\to}\0 A\0; &\qquad A \0\0 \overset{1}{\to}\0 A\0; \\
\0 A\0 \overset{1}{\to}\0\0 A; &\qquad \0 A\0 \overset{1}{\to} A\0\0;\\
%1
\0 A\0 \overset{\omega}{\to} A\0 A; &\qquad  A\0 A \overset{1}{\to}\0 A\0; \\
%5
\0 A A \overset{2}{\to} A A\0; &\qquad  A A\0 \overset{2}{\to}\0 A A;\\
%7
\0 A\0 \overset{\omega}{\to} A A A; &\qquad  A A A \overset{\frac{1}{\omega}}{\to}\0 A\0; \\
%8
\0 A A \overset{\omega(2-\omega)}{\to} A A A; &\qquad  A A\0 \overset{\omega(2-\omega)}{\to} A A A; \\
 A A A \overset{2-\omega}{\to} \0 A A; &\qquad  A A A \overset{2-\omega}{\to} A A\0;\\
%9
\omega\le 2; &\\
\rho_2=\frac{\omega}{\omega+1}; \qquad \delta_1=\delta_4=0;&\qquad \delta_2=\frac{2}{1-\rho_2}; \qquad \delta_3=2
\end{split}
\end{equation}
For $\omega=2$ some of the rates vanish.

\item
\label{3PWW:modelJ}
\begin{equation}
\begin{split}
A\0 \overset{1}{\rightarrow} \0 A; & \qquad \0 A \overset{1}{\rightarrow} A \0;\\
%DIFF
\0 A A \overset{1}{\to} A A\0; &\qquad  A A\0 \overset{1}{\to}\0 A A;\\ 
%7
\0 A\0 \overset{\omega}{\to} A A A; &\qquad  A A A \overset{\frac{1}{\omega}}{\to}\0 A\0; \\
%8
\0 A A \overset{\omega-1}{\to} A A A; &\qquad  A A\0 \overset{\omega-1}{\to} A A A; \\
 A A A \overset{\frac{\omega-1}{\omega}}{\to} \0 A A; &\qquad  A A A \overset{\frac{\omega-1}{\omega}}{\to} A A\0;\\
%9
\omega\ge 1; &\\
\rho_2=\frac{\omega}{\omega+1}; \qquad \delta_1=\delta_4=0;&\qquad \delta_2=\frac{1}{1-\rho_2}; \qquad \delta_3=1
\end{split}
\end{equation}
For $\omega=1$ some of the rates vanish.

\item
\begin{equation}
\begin{split}
\0\0 A \overset{1}{\to}\0 A A; &\qquad \0 A A \overset{1}{\to}\0\0 A; \\
A\0 \0 \overset{1}{\to}A A \0; &\qquad  A A\0 \overset{1}{\to} A\0\0;\\
%2
\0\0 A \overset{1}{\to} A\0\0; &\qquad  A\0\0 \overset{1}{\to}\0\0 A;\\
\0\0 A \overset{1}{\to} A\0 A; &\qquad  A\0\0 \overset{1}{\to} A\0 A; \\
 A\0 A \overset{1}{\to}\0\0 A; &\qquad  A\0 A \overset{1}{\to} A\0\0;\\
\0\0 A \overset{1}{\to} A A\0; &\qquad \0 A A \overset{1}{\to} A\0\0;\\
A\0 \0 \overset{1}{\to}\0 A A; &\qquad  A A\0 \overset{1}{\to}\0 A A;\\
\0\0 A \overset{1}{\to} A A A; &\qquad  A\0\0 \overset{1}{\to} A A A; \\
 A A A \overset{1}{\to}\0\0 A; &\qquad  A A A \overset{1}{\to} A\0\0;\\
%3
\0 A\0 \overset{1}{\to} A\0 A; &\qquad  A\0 A \overset{1}{\to}\0 A\0; \\
%5
\rho_2=\frac{1}{2}; \qquad \delta_1=4;\qquad \delta_2=2;&\qquad \delta_3=1; \qquad \delta_4=1
\end{split}
\end{equation}

\item~\begin{equation}
\begin{split}
\0\0 A \overset{1}{\to} A\0\0; &\qquad  A\0\0 \overset{1}{\to}\0\0 A; \\
\0\0 A \overset{1}{\to} A\0 A; &\qquad  A\0\0 \overset{1}{\to} A\0 A; \\
 A\0 A \overset{1}{\to}\0\0 A; &\qquad  A\0 A \overset{1}{\to} A\0\0;\\
\0\0 A \overset{1}{\to} A A\0; &\qquad \0 A A \overset{1}{\to} A\0\0; \\
A\0 \0 \overset{1}{\to}\0 A A; &\qquad  A A\0 \overset{1}{\to}\0 A A;\\
\0\0 A \overset{1}{\to} A A A; &\qquad  A\0\0 \overset{1}{\to} A A A; \\
 A A A \overset{1}{\to}\0\0 A; &\qquad  A A A \overset{1}{\to} A\0\0;\\
%3
\0 A\0 \overset{1}{\to}\0 A A; &\qquad \0 A\0 \overset{1}{\to} A A\0; \\
\0 A A \overset{1}{\to}\0 A\0; &\qquad  A A\0 \overset{1}{\to}\0 A\0;\\
%4
\0 A\0 \overset{1}{\to} A\0 A; &\qquad  A\0 A \overset{1}{\to}\0 A\0;\\
%5
\0 A A \overset{1}{\to} A A A; &\qquad  A A\0 \overset{1}{\to} A A A; \\
 A A A \overset{1}{\to} \0 A A; &\qquad  A A A \overset{1}{\to} A A\0;\\
%9
\rho_2=\frac{1}{2}; \qquad \delta_1=4; \qquad\delta_2=2;&\qquad \delta_3=1; \qquad \delta_4=1
\end{split}
\end{equation}

\item
\begin{equation}
\begin{split}
\0\0 A \overset{1}{\to} A\0\0; &\qquad  A\0\0 \overset{1}{\to}\0\0 A; \\
\0\0 A \overset{1}{\to} A\0 A; &\qquad  A\0\0 \overset{1}{\to} A\0 A; \\
 A\0 A \overset{1}{\to}\0\0 A; &\qquad  A\0 A \overset{1}{\to} A\0\0;\\
\0\0 A \overset{1}{\to} A A\0; &\qquad \0 A A \overset{1}{\to} A\0\0; \\
A\0 \0 \overset{1}{\to}\0 A A; &\qquad  A A\0 \overset{1}{\to}\0 A A;\\
\0\0 A \overset{1}{\to} A A A; &\qquad  A\0\0 \overset{1}{\to} A A A; \\
 A A A \overset{1}{\to}\0\0 A; &\qquad  A A A \overset{1}{\to} A\0\0;\\
%3
\0 A\0 \overset{1}{\to}\0 A A; &\qquad \0 A\0 \overset{1}{\to} A A\0; \\
\0 A A \overset{1}{\to}\0 A\0; &\qquad  A A\0 \overset{1}{\to}\0 A\0;\\
%4
\0 A A \overset{1}{\to}\0 A A; &\qquad \0 A\0 \overset{1}{\to} A A\0; \\
 \0 A A \overset{1}{\to} \0\0 A; &\qquad  A A A \overset{1}{\to} \0 A\0;\\
%6
\0 A\0 \overset{1}{\to} A A A; &\qquad  A A A \overset{1}{\to}\0 A\0;\\
%8
\rho_2=\frac{1}{2}; \qquad \delta_1=4; \qquad\delta_2=2;&\qquad \delta_3=1; \qquad \delta_4=1
\end{split}
\end{equation}

\item
\begin{equation}
\begin{split}
\0\0 A \overset{1}{\to} A\0\0; &\qquad  A\0\0 \overset{1}{\to}\0\0 A;\\
\0\0 A \overset{1}{\to} A\0 A; &\qquad  A\0\0 \overset{1}{\to} A\0 A; \\
 A\0 A \overset{1}{\to}\0\0 A; &\qquad  A\0 A \overset{1}{\to} A\0\0;\\
\0\0 A \overset{1}{\to} A A\0; &\qquad \0 A A \overset{1}{\to} A\0\0; \\
A\0 \0 \overset{1}{\to}\0 A A; &\qquad  A A\0 \overset{1}{\to}\0 A A;\\
\0\0 A \overset{1}{\to} A A A; &\qquad  A\0\0 \overset{1}{\to} A A A; \\
 A A A \overset{1}{\to}\0\0 A; &\qquad  A A A \overset{1}{\to} A\0\0;\\
%3
\0 A\0 \overset{1}{\to} A\0 A; &\qquad  A\0 A \overset{1}{\to}\0 A\0; \\
%5
\0 A A \overset{1}{\to} A A\0; &\qquad  A A\0 \overset{1}{\to}\0 A A;\\
%7
\0 A\0 \overset{1}{\to} A A A; &\qquad  A A A \overset{1}{\to}\0 A\0; \\
%8
\rho_2=\frac{1}{2}; \qquad \delta_1=4;\qquad \delta_2=2;&\qquad \delta_3=1; \qquad \delta_4=1
\end{split}
\end{equation}

\end{enumerate}
}

\end{document}